\documentclass[a4paper,10pt,onecolumn,oneside,notitlepage,openbib]{article}
\usepackage{amssymb}
\usepackage{amsmath}

\author{
  Fumitaka Yura\thanks{Imai quantum computing and information project,
    ERATO, JST, Daini Hongo White Bldg. 201, 5-28-3 Hongo, Bunkyo,
    Tokyo 133-0033, Japan} \ 
   and Tetsuji Tokihiro\thanks{Graduate school of Mathematical Sciences,
      University of Tokyo, 3-8-1 Komaba, Tokyo 153-8914, Japan}}
\title{On a Periodic Soliton Cellular Automaton}
\date{}

\usepackage{graphics}
\usepackage{theorem}
\theoremstyle{break}
\newtheorem{Theorem}{Theorem}
\newtheorem{Proof}{Proof}
\newtheorem{Proposition}{Proposition}
\newtheorem{Lemma}{Lemma}
\newtheorem{Corollary}{Corollary}
\newtheorem{Example}{Example}

\def\ket#1{|#1\rangle}
\def\C{{\mathbb C}}
\def\Z{{\mathbb Z}}
\def\F{{\mathbb F}}
\def\mapto{\rightarrow}
\def\ball{\hspace*{-0.25em}\raisebox{-0.7ex}{\huge $\bullet$}}

\begin{document}
\maketitle

\begin{abstract}

We propose a box and ball system with a periodic boundary condition (pBBS).
The time evolution rule of the pBBS is represented as a Boolean recurrence formula, an inverse ultradiscretization of which is shown to be equivalent
with the algorithm of the calculus for the $2N$th root.
The relations to the pBBS of the combinatorial $R$ matrix of 
${U'}_q(A_N^{(1)})$ are also discussed.

\end{abstract}

\section{Preface}

In many physical phenomena, physical quantities and time-space variables are continuous.
Discretized models, however, are used sometimes for simplification and/or 
speedup of computation.
In discretizing process, it is preferable to keep the mathematical structure
(symmetry, conserved quantities, {\it etc.}) of the original
 continuous systems.
For integrable systems, {\it i.e.} the systems which exhibit solitonic natures
and are described by integrable nonlinear partial differential equations, 
several {\it effective} methods of discretization have been established and
many discrete integrable systems have been proposed\cite{HTI}.
An effective method to create a cellular automaton (CA) with solitonic natures
is the ultradiscretization\cite{TTMS} which is a limiting procedure to 
discretize the dependent variables of a partial difference equation.

In this paper, we consider an integrable cellular automaton called 
box and ball system (BBS)\cite{T, TNS}.
The BBS has been defined as the time evolution of
a finite number of balls moving in one
dimensional array of {\it infinite} number of boxes.
We extend the time evolution rule of the BBS so that the BBS can be defined 
when we impose a periodic boundary condition.
We show that the time evolution rule of the periodic BBS (pBBS)
is given by a Boolean formula,
which turns out to be an algorithm to calculate the $2N$th root of a 
given number through inverse ultradiscretization.
It is well known that many of the discrete
integrable systems are equivalent to good algorithms such as the 
diagonalization of matrix (QR algorithm, {\it etc.}),
 convergence acceleration
algorithm ($\varepsilon$ algorithm, {\it etc.}), Karmarkar algorithm, 
BCH-Goppa decoding.
Our Boolean formula is an example of such correspondence between
discrete integrable systems and algorithms.

In section 2, we introduce pBBS and explain its time evolution rule.
In section 3, we obtain recurrence formulae on Boolean algebra for the
time evolution rule of the simplest BBS with box capacity one.
The correspondence of the Boolean formula with the algorithm for the computation of the $2N$th root is shown.
In section 4, we establish a relation of the pBBS to the combinatorial 
$R$ matrix model and extend the time evolution rule to the general pBBSs
with various box capacities.
Section 5 is devoted to the concluding remarks.

\section{Box and ball system}

\subsection{Infinite BBS}

The BBS is a reinterpretation and extension of the filter-type 
CA proposed by Takahashi and Satsuma\cite{TTMS, TS}.

Let us consider a one dimensional array of infinite number of boxes.
At the initial time $t=0$, all but finite number of the boxes are empty,
and each of the rest boxes contains one ball such as:
\begin{center}
\begin{tabular}{p{1.0em}*{13}{| p{0.5em} }p{0em}}
\hline
$\cdots$ & & & \ball & \ball & & \ball& \ball & & \ball & & & & $\cdots$ &\\
\hline
\end{tabular}
\end{center}

\begin{figure}[!b]
  \begin{center}
     \scalebox{0.7}{\includegraphics{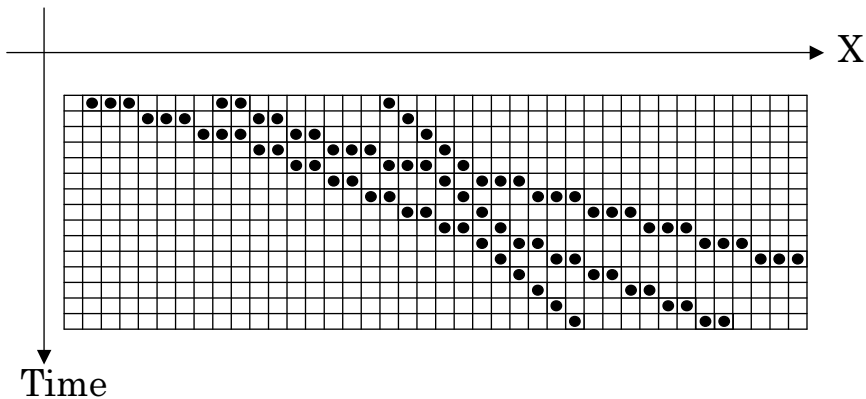}}
  \end{center}
  \caption{An example of the time evolution of BBS.}
  \label{fig:exBBS}
\end{figure}

The time evolution rule of this system from time $t$ to $t+1$ is 
given as follows.
\begin{enumerate}
  \item Move every ball only once.
  \item Move the leftmost ball to its nearest right empty box.
  \item Move the leftmost ball of the rest ball to its nearest right empty
  box.
  \item Repeat the above procedure until all the balls are moved.
\end{enumerate}
An example of the time evolution is shown in Fig.~\ref{fig:exBBS}.
In the example, we see the solitonic behavior of balls.
This is a general behavior of the BBS and, starting from an arbitrary 
initial state, we can prove that the state asymptotically evolves into 
the state that consists of only freely moving solitons.
It is also known that the BBS has infinite number of conserved 
quantities\cite{TTS}.
By replacing an empty box by "0" and a filled box by "1", we regard
the BBS as a dynamical system of "01" sequence, namely, a CA.
The relation between BBS and (classical) integrable systems, {\it i.e.}
integrable partial differential and/or difference equations, was established
by the notion of ultradiscretization (UD)\cite{TTMS}.
In the approach of UD, the BBS is constructed through a limit of
the partial difference 
equation, which is obtained by reduction of the discrete KP equation 
(Hirota-Miwa equation).
Since the discrete KP equation is equivalent to the generating formula of 
the KP hierarchy, the BBS naturally inherits its integrability and shows
solitonic natures.
In this sense, the BBS is called an integrable CA.

\subsection{Periodic BBS (pBBS)}

\begin{figure}[!b]
  \begin{center}
    \scalebox{0.8}{\includegraphics{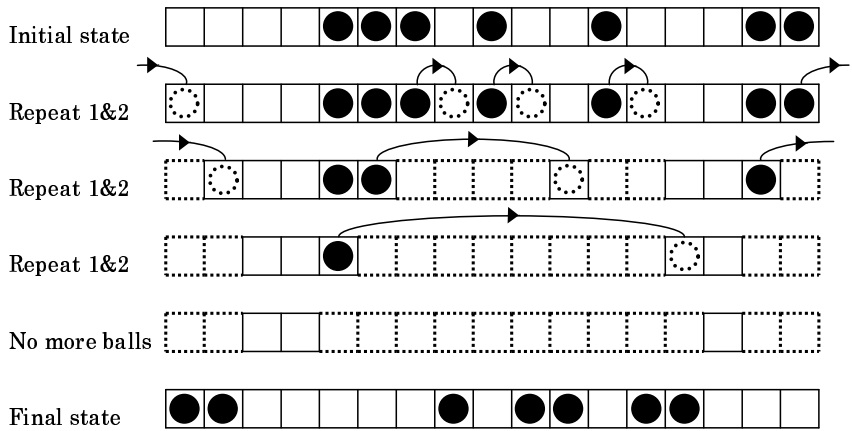}}
  \end{center}
  \caption{An example of the modified rule for pBBS.}
  \label{fig:pBBSrule}
\end{figure}

In this subsection, we extend the original BBS to that 
with a periodic boundary condition.
Let the BBS consist of $N$ boxes.
To impose a periodic boundary condition, we assume that
the $N$th box is the adjacent box to the first box. (We may imagine
that the boxes are arranged in a circle.)
We also assume that the box capacity is one for all the boxes. 
Since the evolution rule of the BBS requires the definition of
 the {\it leftmost} ball in the BBS,
 we cannot apply the original evolution rule directly 
to a periodic system.
Instead, we consider the following time evolution rule:
\begin{enumerate}
  \item Move all the balls to their conterminous right boxes 
        if the boxes are empty.
  \item Forget the boxes to which and from which the balls were moved in
        the first step, and move all of the rest balls to their
        conterminous right boxes if they are empty.
  \item Repeat the above procedure until all the balls are moved.
\end{enumerate}
An example of the movement of the balls by this rule is illustrated 
in Fig.~\ref{fig:pBBSrule}.

\begin{figure}[!b]
  \begin{center}
    \scalebox{0.5}{\includegraphics{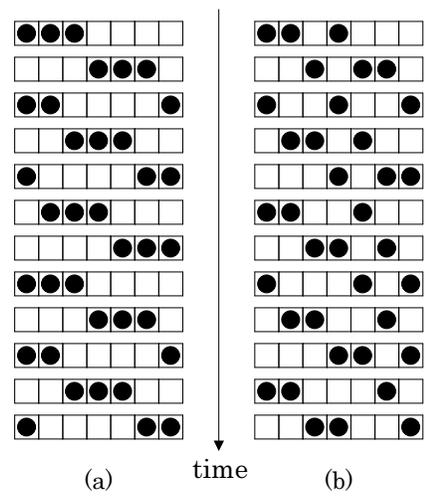}}
  \end{center}
  \caption{Examples of pBBS ($N=7$).}
  \label{fig:N7}
\end{figure}

We see that this rule is equivalent to the original time evolution rule when we apply it to the BBS with an infinite array of the boxes.
In this rule, the number of balls which move at each stage does not change in time evolution.
We will denote by a periodic BBS (pBBS) the BBS with a periodic boundary condition which evolves in this rule.
\begin{Example}[pBBS ($N=7$)]
Time evolution patterns of pBBS are shown in Fig.~\ref{fig:N7}.
(a) The initial state is "1110000". The time evolution pattern has the 
fundamental cycle 7.
(b) The initial state "1101000" has the fundamental cycle 21.
\end{Example}

\begin{figure}[htbp]
  \begin{center}
    \scalebox{0.7}{\includegraphics{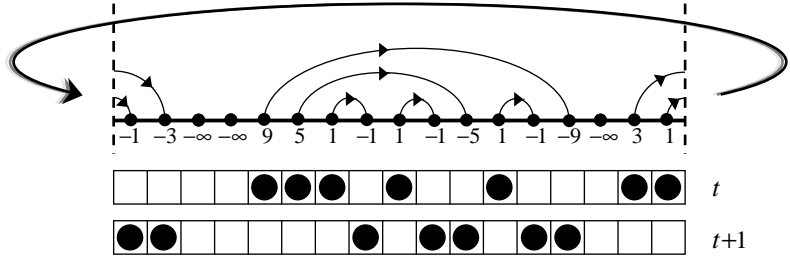}}
  \end{center}
  \caption{A diagram of time evolution and 
  corresponding time evolution of the pBBS.
  The integers below the nodes are the indices $\gamma_t(n)$.
  }
  \label{fig:figdia}
\end{figure}

The time evolution of pBBS is represented by a diagram with $N$ nodes 
and $M$ lines each of which connects two of the distinct nodes as shown
in Fig.~\ref{fig:figdia}.

Each node corresponds to a box of the pBBS and a line is drawn
 from the box which contains a ball to the box to which the ball is moved.
From the time evolution rule of the pBBS, we find the following properties of the diagram:
\begin{itemize}
\item There is no intersection of the lines.
\item There is no line over the nodes which are not connected by lines.
\item If we remove the unconnected nodes, the diagram is divided
      into a set of {\it blocks} with even number of nodes 
      which are connected by lines pairwisely.
\end{itemize}
For convenience of explanation in the subsequent sections, we define indices $\{\gamma_t(n)\}_{n=1}^{N}$ of the nodes in the diagram at time step $t$:
\begin{itemize}
\item If a ball in the $n$th box is moved to the $n+k$th (modulo $N$) box
      at $t+1$, $\gamma_t(n)=k (\in \Z_{>0})$.
\item If a ball is moved to the $n$th box from the $n-k$th (modulo $N$) box 
      at $t+1$, $\gamma_t(n)=-k$. 
\item Otherwise, $\gamma_t(n)=-\infty$
\end{itemize}
In the example shown in Fig.~\ref{fig:figdia}, the indices $\{\gamma_t(n)\}$
$=$ $\{$ $-1$, $-3$, $-\infty$, $-\infty$, $9$, $5$, $1$, $-1$, $1$,
 $-1$, $-5$, $1$, 
$-1$, $-9$, $-\infty$, $3$, $1$ $\}$.
The following Proposition is a direct consequence of the properties of the
diagram.

\begin{Proposition}
\label{Prop.1}
Let $k$ and $l$ ( $l < k$ ) be positive integers. Then
\begin{itemize}
\item[(a)] If $\gamma_t(n)=k$, then $\gamma_t(n+k)=-k$, and
$\ -(k-2) \le \gamma_t(n+j) \le k-2 \ $ for $\ 1 \le j \le k-1$.
\item[(b)]
Furthermore if there exists $j$ such that $1 \le j \le k-1$ and
$\gamma_t(n+j)=-l$, then $\ l < j\ $ and 
$\ \gamma_t(n+j-l)=l$.
\end{itemize}
\end{Proposition}

In Proposition \ref{Prop.1}, the indices are understood with the 
convention of {\it modulo $N$}, {\it i.e.} $\gamma_t(n) \equiv \gamma_t(n+N)$.
Hereafter, we often use this convention.

There are several equivalent evolution rules for the pBBS.
For example,
\begin{enumerate}
  \item At each filled box, create a copy of the ball.
  \item Move all the copies once according to the following rule.
  \item Choose one of the copies and move it to its nearest right empty box.
  \item Choose one of the rest copies and move it to its nearest right empty box.
  \item Repeat the above procedure until all the copies are moved.
  \item Delete all the original balls.
\end{enumerate}
It is not difficult to prove that the result does not depend on the choice of 
the copies at each stage, and that this rule gives the same time evolution 
pattern of the previous rule.
An advantage of this rule is that it is straightforward to 
extend the rule to the BBS with many kinds of balls 
and various box capacities.
We can also use the combinatorial $R$ matrix 
of ${U'}_q(A_{N-1}^{(1)})$ to give an equivalent evolution rule, the detail
of which is presented in section 4.

\section{Recurrence equations and corresponding algorithm}

\subsection{Boolean formulae of pBBS}

We show that pBBS introduced above can be formulated by Boolean algebra.
Let $N$ be the number of the boxes.
The space of the states of pBBS is naturally regarded as ${\F_2}^N$.
We denote a state $X(t)$ of the pBBS at time $t$ by $X(t)=(x_1(t),x_2(t),
\ldots, x_N(t)) \in {\F_2}^N$,
where $x_i(t)=0$ if the $i$th box is empty and $x_i(t)=1$ if it is filled.
Let $\wedge $, $\vee $, $\oplus $, be AND, OR and XOR respectively.
These Boolean operators are realized in ${\F_2}^N$ as the map:
${\F_2}^N \times {\F_2}^N \to {\F_2}^N$.
For $X=(x_1, x_2, \ldots, x_N)$, $Y=(y_1, y_2, \ldots, y_N)$,
they are defined as
\begin{eqnarray*}
(X \wedge Y)_i &:= x_i \wedge y_i &\equiv x_iy_i \\
(X \vee Y)_i &:= x_i \vee y_i &\equiv x_iy_i + x_i + y_i \\
(X \oplus Y)_i &:= x_i \oplus y_i &\equiv x_i + y_i.
\end{eqnarray*}
We also define rotate shift to the right $S$:
\[
  S X = (x_N, x_1, x_2,\ldots , x_{N-1}).
\]
Next theorem gives an expression of $T$
\[
  T : X(t) \mapsto X(t+1)
\]
in terms of these Boolean operators.

\begin{Theorem}
\label{Theorem1}
Suppose that $X(t) \in {\F_2}^N$ is the state of pBBS
at time step $t$. We consider the following recurrence equations:
\begin{eqnarray}
  && \; A^{(0)}=X(t), \ B^{(0)}=S X(t),\\
  &&\left\{
    \begin{array}{l}
      A^{(n+1)} := A^{(n)} \vee B^{(n)} \\
      B^{(n+1)} := S (A^{(n)} \wedge B^{(n)}) \\
    \end{array}
  \right.   \quad (n=0,1,2,\ldots) .
  \label{eq:req1}
\end{eqnarray}
Then, 
\begin{equation}
X(t+1) = A^{(N)} \oplus X(t),\ \mbox{\rm and} 
\  B^{(N)}=\mbox{\boldmath{$0$}},
\label{eq:reqfin}
\end{equation}
where $\mbox{\boldmath{$0$}}:=(0,0,\ldots,0)$.
\end{Theorem}

\begin{Proof}
We define $D^{(i)}=(d_1^{(i)}, d_2^{(i)}, \ldots, d_N^{(i)})
 \in {\F_2}^N$ $(i=1, 2, \ldots, N)$ as follows.
If the $l$th box contains a ball at time step $t$ and it moves to 
the $l+i$th box at $t+1$, then $d_l^{(i)}=1$, otherwise $d_l^{(i)}=0$.
If we use the indices $\gamma_t(l)$ defined in the previous section, 
the definition of $D^{(i)}$ is rewritten as \\ 
\begin{center}
$d_l^{(i)}=1$ if and only if $\gamma_t(l)=i$.
\end{center}

Clearly, these $D^{(i)}$ give the unique decomposition of the initial state 
$X(t)$ and the final state $X(t+1)$ as
\begin{eqnarray*}
  &&X(t)=\displaystyle \bigoplus_{i=1}^{N} D^{(i)}, \ \  
   D^{(i)} \wedge D^{(j)}=\mbox{\boldmath{$0$}}  \ (i\neq j) ,\\
  &&T X(t)=\displaystyle \bigoplus_{i=1}^{N} S^i D^{(i)}, \ \ 
  S^i D^{(i)} \wedge S^j D^{(j)} = \mbox{\boldmath{$0$}} \ (i\neq j),
\end{eqnarray*}
where $\bigoplus_{i=1}^N D^{(i)}:= D^{(1)} \oplus D^{(2)} \oplus 
\ldots \oplus D^{(N)}$.

To prove the theorem, it suffices to prove the following formulae
\begin{eqnarray}
A^{(n)}&= &X(t) \oplus \bigoplus_{i=1}^{n} S^i D^{(i)} ,
\label{eq:An}\\
S^{-n-1}B^{(n)}&= &X(t) \oplus \bigoplus_{i=1}^{n} D^{(i)} .
\label{eq:Bn}
\end{eqnarray}
Indeed, if (\ref{eq:An}) and (\ref{eq:Bn}) hold, we have
\begin{eqnarray*}
A^{(N)} \oplus X(t) &= &X(t) \oplus \bigoplus_{i=1}^{N} S^i D^{(i)} 
\oplus X(t)\\
& =& \bigoplus_{i=1}^{N} S^i D^{(i)} \\
& =& X(t+1), \\
B^{(N)} & =& S^{N+1} \left( X(t) \oplus \bigoplus_{i=1}^{N} D^{(i)} \right) \\
        & =& S^{N+1} \left( X(t) \oplus X(t) \right)\\
        & =& \mbox{\boldmath{$0$}}.
\end{eqnarray*}
We prove (\ref{eq:An}) and (\ref{eq:Bn}) by induction.

For $n=1$,
\begin{eqnarray*}
A^{(1)} &= &A^{(0)} \vee B^{(0)}\\
        &= &X(t) \vee S X(t)\\
        &= &X(t) \vee S D^{(1)} \vee \bigoplus_{i=2}^{N} S D^{(i)} .
\end{eqnarray*}
Suppose that $ \bigoplus_{i=2}^{N} S D^{(i)} \vee X(t) \ne X(t)$, 
 there exists $D^{(q)}$ $(q \ge 2)$ such that its $n'$ component
$d_{n'}^{(q)}=1$ and $\gamma_t(n'+1) \le -1$.
However, by the definition of $D^{(q)}$, $d_{n'}^{(q)}=1$ implies
$\gamma_t(n')=q (\ge 2)$.
From Proposition \ref{Prop.1} (a), $-q+2 \le \gamma_t(n'+1) \le q-2$
and we find $-q+2 \le \gamma_t(n'+1) \le -1$,
 which contradicts Proposition \ref{Prop.1} (b). 
Hence $ \bigoplus_{i=2}^{N} S D^{(i)} \vee X(t) = X(t)$.
Since $S D^{(1)} \wedge X(t) = \mbox{\boldmath{$0$}}$ follows from
$TX(t) \wedge X(t)=\mbox{\boldmath{$0$}}$, we find
$$
A^{(1)}=X(t) \oplus S D^{(1)}.
$$
Similarly, from the relations $X(t) \wedge \bigoplus_{i=2}^N S D^{(i)} 
= \bigoplus_{i=2}^N S D^{(i)}$ and $S D^{(1)} \wedge X(t) =
\mbox{\boldmath{$0$}}$,
we have
\begin{eqnarray*}
S^{-2}B^{(1)} &= &S^{-1} \left( X(t) \wedge S X(t) \right)\\
              &= &S^{-1} \left( X(t) \wedge \bigoplus_{i=1}^N
               S D^{(i)} \right)\\
              &= &S^{-1} \left(X(t) \wedge S D^{(1)}\right) \oplus
              S^{-1} \left( X(t) \wedge \bigoplus_{i=2}^N S D^{(i)}\right) \\
              &= &\left( S^{-1} \mbox{\boldmath{$0$}}\right) \oplus
              S^{-1}\left( \bigoplus_{i=2}^N S D^{(i)} \right)\\
              &= &\bigoplus_{i=2}^N D^{(i)}\\
              &= & X(t) \oplus D^{(1)}.
\end{eqnarray*}
Hence (\ref{eq:An}) and (\ref{eq:Bn}) hold for $n=1$.

Assume that (\ref{eq:An}) and (\ref{eq:Bn}) is true for $n=k$, then
\begin{eqnarray*}
A^{(k+1)} &= &\left( X(t) \oplus \bigoplus_{i=1}^{k} S^i D^{(i)}\right)
             \vee \left( S^{k+1} X(t) \oplus \bigoplus_{i=1}^{k} S^{k+1}D^{(i)}
             \right)\\
          &= &\left( X(t) \oplus \bigoplus_{i=1}^{k} S^i D^{(i)}\right)
             \vee \left(\bigoplus_{i=k+1}^{N} S^{k+1}D^{(i)}
             \right).
\end{eqnarray*}
By the definition of $D^{(k+1)}$, we have 
$$
\left( X(t) \oplus \bigoplus_{i=1}^{k} S^i D^{(i)}\right) \vee S^{k+1}D^{(k+1)}
= X(t) \oplus \bigoplus_{i=1}^{k+1} S^i D^{(i)}.
$$
Suppose that
$$
\left( X(t) \oplus \bigoplus_{i=1}^{k} S^i D^{(i)}\right) \vee \bigoplus_{i=k+2}^{N} S^{k+1} D^{(i)} \ne X(t) \oplus \bigoplus_{i=1}^{k} S^i D^{(i)}.
$$
Noticing the fact that, for $X(t) \oplus \bigoplus_{i=1}^{k} 
S^i D^{(i)}$ $=$
$(a_1^{(k)}, a_2^{(k)}, \ldots, a_N^{(k)})$, $a_j^{(k)}=1$
if and only if $-k \le \gamma_t(j)$,
we find that there is at least one $D^{(q)}$ $(q \ge k+2)$ one of whose
components satisfies $d_{n'}^{(j)}=1$ and $\gamma_t(n'+k+1) \le -k-1$.
However, by the definition of $D^{(q)}$, $d_{n'}^{(q)}=1$ implies
$\gamma_t(n')=q (\ge k+2)$.
From Proposition \ref{Prop.1} (a), $-q+2 \le \gamma_t(n'+k+1) \le q-2$
and we find $-q+2 \le \gamma_t(n'+k+1) \le -k-1$,
 which contradicts Proposition \ref{Prop.1} (b). 
Thus, we have
$$
A^{(k+1)}= X(t) \oplus \bigoplus_{i=1}^{k+1} S^i D^{(i)}.
$$
In a similar manner, we also obtain
$$
S^{-k-2}B^{(k+1)} = X(t) \oplus \bigoplus_{i=1}^{k+1} D^{(i)}.
$$
Hence, from the assumption of induction, the formulae (\ref{eq:An}) 
and (\ref{eq:Bn}) are proved to hold for $0 \le n \le N-1$, which completes
the proof of the thorem.
\end{Proof}

This recurrence equation (\ref{eq:reqfin}) is expressed with only
three operations, AND, OR and SHIFT, and has a simple form.
The SHIFT operator introduces the right-and-left symmetry breaking 
that comes from the definition of the direction of the movement of balls.

Since $D^{(2m)}=\mbox{\boldmath{$0$}}$, we immediately 
obtain the following Corollary:

\begin{Corollary}

Suppose that $X(t) \in {\F_2}^N$ is given as the state at time $t$.
Then the state at next time $X(t+1) = T X(t)$ is calculated by recurrence equation as follows.
 
\begin{eqnarray}
&& \; A^{(0)}:=X(t), \ B^{(0)}:=SX(t) \\
&&\left\{
    \begin{array}{l}
      A^{(n+1)} := A^{(n)} \vee B^{(n)} \\
      B^{(n+1)} := S^2(A^{(n)} \wedge B^{(n)}) \\
    \end{array}
  \right. ,
  \label{eq:req2}
\end{eqnarray}
and 
\begin{equation}
  X(t+1) = A^{(\left\lfloor N/2 \right\rfloor)} \oplus X(t),\ 
  B^{(\left\lfloor N/2 \right\rfloor)}=\mbox{\boldmath{$0$}} .
  \label{eq:reqfin2}
\end{equation}
\end{Corollary}

\subsection{pBBS and numerical algorithm}

The formulae for time evolution of the pBBS (\ref{eq:req1})
 have simple and symmetric form,
and we expect that they have some relations to a
good algorithm. 
In this subsection, we show that they have indeed the same structure as that of
the algorithm to compute $N$th root of a given number.
Henceforth, let the truth value "0(false)" and "1(true)" be equivalent to the integer $0 \in \Z$ and $1 \in \Z$.
Then we can replace $\wedge$ and $\vee$ with $\min$ and $\max$ as
\[
  \left\{
    \begin{array}{lll}
      x \wedge y & \iff & \min \left[x, y \right] \\
      x \vee y   & \iff & \max \left[x, y \right] \\
    \end{array}
  \right. .
  \label{eq:BtoZ}
\]

Following the notation in the previous section, we define that 
$\max$ and $\min$ act on $\Z^N$ bitwisely.
Then, Eq.~(\ref{eq:req1}) can be rewritten by the equation of integers as
\begin{equation}
  \left\{
    \begin{array}{l}
      A^{(n+1)} = \max \left[ A^{(n)}, B^{(n)} \right] \\
      B^{(n+1)} = S \min \left[ A^{(n)}, B^{(n)} \right]
    \end{array}
  \right. .
\label{eq:minmax}
\end{equation}
We construct the difference equations corresponding to (\ref{eq:minmax})
by means of inverse ultradiscretization\cite{TTMS}.
Noticing the identity:
\[
  \max \left[x, y \right] = \lim_{\epsilon \to +0}
    \epsilon \log \left( e^{x/\epsilon} + e^{y/\epsilon} \right)
    \quad (x, y \in \mathbb{R}),
\]
and $\min\left[x, y\right] = -\max\left[-x, -y\right]$,
we think of the difference equations:
\begin{equation}
  \left\{
  \begin{array}{l}
    a_i^{(n+1)} = \left\{\displaystyle a_i^{(n)}+b_i^{(n)}\right\}/2 \\
    b_i^{(n+1)} = 2 \left\{
      \left( a_{i-1}^{(n)}\right)^{-1}+\left( b_{i-1}^{(n)}\right)^{-1}
      \right\}^{-1} \\
  \end{array}
  \right. \;\; (1 \le i \le N).
  \label{eq:req4}
\end{equation}
The relation between (\ref{eq:minmax}) and (\ref{eq:req4}) is obvious.
When we replace $a_i^{(n)}$ and $b_i^{(n)}$ with 
$\displaystyle e^{(A^{(n)})_i/\epsilon}$ and $\displaystyle
 e^{(B^{(n)})_i/\epsilon}$ respectively, and take a limit $\epsilon \to +0$, 
we obtain (\ref{eq:minmax}) from (\ref{eq:req4}).
The factor $2$ in (\ref{eq:req4}) is so chosen that the recurrence formulae 
do not diverge at $n \to \infty$.

When we disregard the space coordinates $i$ in (\ref{eq:req4}),
 or consider the case $N=1$, we have recurrence formulae
\begin{equation}
  \left\{
  \begin{array}{l}
    a^{(n+1)} = \frac{\displaystyle a^{(n)}+b^{(n)}}{\displaystyle 2} \\
    b^{(n+1)} = \frac{\displaystyle 2a^{(n)}b^{(n)}}
                     {\displaystyle a^{(n)}+b^{(n)}} \\
  \end{array}
  \right. ,
  \label{eq:AH}
\end{equation}
which is the well-known arithmetic-harmonic mean algorithm and we have
\[
  \lim_{n\to\infty} a^{(n)} = \lim_{n\to\infty} b^{(n)} = \sqrt{a^{(0)}b^{(0)}} .
\]
The recurrence formulae (\ref{eq:req4}) for general $N$ is also considered 
as a numerical algorithm to calculate the $2N$th root of a given number.
To see this, first we note that
(\ref{eq:req4}) has a conserved quantity $C$ with respect to
the step $n$,
\begin{equation}
 C^{(n)} := \displaystyle \prod_{i=1}^{N} a_i^{(n)} b_i^{(n)}= C^{(n-1)} 
 = \cdots = C^{(0)} = 
    \displaystyle \prod_{i=1}^{N} a_i^{(0)} b_i^{(0)} \equiv C
  \label{eq:conserve1} ,
\end{equation}
where $\left\{ a_i^{(0)}, b_i^{(0)} \right\}$ are the initial values.
Then we can show the following Proposition:

\begin{Proposition}
If all the initial values $\{ a_i^{(0)}, b_i^{(0)} \}$ are positive,
then they converge to the same value
\[
  \lim_{n\to\infty} a_k^{(n)} = \lim_{n\to\infty} b_k^{(n)}
     = \sqrt[2N]{\prod_{i=1}^{N} a_i^{(0)} b_i^{(0)} }
     = \sqrt[2N]{C} \mbox{\hspace{1em}(for all $k$)}.
\]
Hence, the recurrence formula of pBBS is regarded as a numerical 
algorithm of $2N$th root.
\label{Prop.2Nroot}
\end{Proposition}
To prove the Proposition, we need a Lemma:

\begin{Lemma}
For $m >0$, $\alpha \ge 0$, and $\varepsilon >0$ which satisfy
\begin{equation}
(2\alpha^2 + 5 \alpha +4) \varepsilon < m, 
\label{m.alpha.Ineq}
\end{equation}
if it holds that
$m-\varepsilon < a_i^{(n+1)} < m+ \alpha \varepsilon$, 
$m-\varepsilon < a_i^{(n)}$ and $m-\varepsilon < b_i^{(n)}$, then
$a_i^{(n)}, b_i^{(n)} < m+(2\alpha+2)\varepsilon$.
Similarly, if it holds that
$m-\varepsilon < b_{i+1}^{(n+1)} < m+ \alpha \varepsilon$, 
$m-\varepsilon < a_i^{(n)}$ and $m-\varepsilon < b_i^{(n)}$, then
$a_i^{(n)}, b_i^{(n)} < m+(2\alpha+2)\varepsilon$.
\label{Lemma.Ineq}
\end{Lemma}

This Lemma is proved from (\ref{eq:req4}) by straightforward calculations.
Now we give the proof of the Proposition \ref{Prop.2Nroot}.

\begin{Proof}
Let $m^{(n)}:=\min_{i=1, \ldots, N}\left[a_i^{(n)}, b_i^{(n)}\right]$ and
$M^{(n)}:=\max_{i=1, \ldots, N}\left[a_i^{(n)}, b_i^{(n)}\right]$. 
Since we have from (\ref{eq:req4})
\begin{eqnarray*}
\min\left[a_i^{(n)}, b_i^{(n)}\right] & \le a_i^{(n+1)} & \le \max\left[a_i^{(n)}, b_i^{(n)}\right]\\
\min\left[a_i^{(n)}, b_i^{(n)}\right] & \le b_{i+1}^{(n+1)} & \le \max\left[a_i^{(n)}, b_i^{(n)}\right],
\end{eqnarray*}
we obtain
\begin{equation}
m^{(n)} \le m^{(n+1)} \le M^{(n+1)} \le M^{(n)} \quad 
\mbox{for $n=0,1,2,\ldots$}.
\label{Inequality.mM}
\end{equation}
From (\ref{Inequality.mM}), we find ${}^{\exists}m,\ m=\lim_{n\to \infty} m^{(n)}$ and ${}^{\exists}M,\ M=\lim_{n\to \infty} M^{(n)}$.
Clearly $m \ge m^{(n)},$ $M \le M^{(n)}$ and $m \le M$.

We will prove $m=M$ by leading contradiction to the assumption $m<M$.
Since $m^{(n)}$ and $M^{(n)}$ converge to $m$ and $M$ respectively, 
for all $\varepsilon >0$, there exists $n_0$ such that 
$0 \le m-m^{(n)}< \varepsilon$ and $0 \le M^{(n)} - M < \varepsilon$ for 
${}^{\forall}n \ge n_0$. We take $\varepsilon = \min\left[ (M-m)/2^{N+1}, \ 
m/2^{2N+4}\right]$. Note that, with this choice, 
the inequality (\ref{m.alpha.Ineq}) holds for $0 \le \alpha \le 2^{N+1}$.
For $n=n_0+N$, there exists $a_j^{(n_0+N)}$ or $b_j^{(n_0+N)}$ which
is equal to $m^{n_0+N}$. When $a_j^{(n_0+N)}=m^{(n_0+N)}$, 
noticing $m-\varepsilon \le m^{(n_0+N)} \le m$, $m-\varepsilon 
\le a_j^{(n_0+N-1)}$ and $m-\varepsilon \le \ b_j^{(n_0+N-1)}$, we obtain
from the Lemma \ref{Lemma.Ineq} that
$m-\varepsilon \le a_j^{(n_0+N-1)}, b_j^{(n_0+N-1)} \le m+2 \varepsilon$.
Similarly, when $b_j^{(n_0+N)}=m^{(n_0+N)}$, we obtain
$m-\varepsilon \le a_{j-1}^{(n_0+N-1)}, b_{j-1}^{(n_0+N-1)} \le m+2 
\varepsilon$.
By repeated use of the Lemma \ref{Lemma.Ineq} in a similar manner, 
we finally obtain
\begin{eqnarray*}
&&{}^{\forall}i \ a_i^{(n_0)} \le m + (2^{N+2}-2)\varepsilon <M \le M^{(n_0)}\\
&&{}^{\forall}i \ b_i^{(n_0)} \le m + (2^{N+2}-2)\varepsilon <M \le M^{(n_0)},
\end{eqnarray*}
which contradicts the definition of $M^{(n_0)}$.
Thus, we have proved $m=M$. Hence all the values converge to the same value $C^{1/2N}$.
\end{Proof}

In the preface of this article, we pointed out that the discrete model has to maintain the mathematical structures of a continuous model in the process of 
ultradiscretization.
When we take ultradiscrete limit of $C$, it is also a conserved 
quantity of the pBBS. In fact, 
\begin{equation}
    C = \displaystyle \prod_{i=1}^{N} a_i^{(0)} b_i^{(0)}
        \hspace*{1em}\stackrel{\mbox{UD}}{\Longrightarrow} \hspace*{1em}
        \sum_{i=1}^{N} \left\{ A^{(0)}_i + B^{(0)}_i \right\}
    \label{eq:conserve2}
\end{equation}
gives the double number of balls in the pBBS.
The number of balls is, to be sure, a conserved quantity of the pBBS.

We can construct other conserved quantities of the recurrence formulae
(\ref{eq:req1}) by means of another inverse ultradiscretization.

From (\ref{eq:req1}), we have
\begin{equation}
\left\{
\begin{array}{l}
A^{(n+1)} \wedge S^{-1}B^{(n+1)}=A^{(n)} \wedge B^{(n)}\\
A^{(n+1)} \vee S^{-1}B^{(n+1)}= A^{(n)} \vee B^{(n)}
\end{array}
\right. .
\end{equation}
When we consider the inverse ultradiscretization of the above equation,
we have
\begin{equation}
\left\{
\begin{array}{l}
a_i^{(n+1)}b_{i+1}^{(n+1)}=a_i^{(n)}b_i^{(n)} \\
a_i^{(n+1)}+b_{i+1}^{(n+1)}=a_i^{(n)}+b_i^{(n)}
\end{array}
\right. .
\label{eq:conserve}
\end{equation}
Thus, for arbitrary $\lambda$, $(\lambda + a_i^{(n)})(\lambda +b_i^{(n)})$
$=$ $(\lambda + a_i^{(n+1)})(\lambda +b_{i+1}^{(n+1)})$ and we find
that 
\begin{equation}
C_n(\lambda):= \prod_{i=1}^{N} (\lambda + a_i^{(n)})(\lambda +b_i^{(n)})
\end{equation}
does not depend on $n$, which means that
any symmetric polynomial with respect to $\{a_i^{(n)}\}$ and 
$\{b_i^{(n)}\}$ does not depend on $n$.
Therefore, the ultradiscrete limit of such symmetric polynomials 
gives $2N$ conserved quantities $S_1, S_2, \ldots, S_{2N}$ 
of (\ref{eq:req1}).
If we denote $B_i^{(n)} \equiv A_{N+i}^{(n)}$, these conserved quantities are
explicitly given as
\begin{eqnarray*}
S_1&:= &\max_i\left[A_i^{(n)} \right] \\
S_2&:= &\max_{i<j}\left[A_i^{(n)}+A_j^{(n)} \right] \\
&&\cdots \\
S_{2N}&:=&\sum_{i=1}^{2N} A_i^{(n)}.
\end{eqnarray*}

\section{pBBS and combinatorial $R$ matrix}

\subsection{pBBS as periodic $A_M^{(1)}$ crystal lattice}

The BBS (of infinite number of boxes) has recently been reformulated 
from the theory of crystal
and the combinatorial $R$ matrix\cite{HIK, NY, HHIKTT}.
In this approach, a time evolution pattern of the BBS corresponds to
a ground state configuration of a solvable lattice which has a symmetry of 
quantum algebra $\left. {U'}_q(A_M^{(1)})\right|_{q \to 0}$.
The Boltzmann weight on every vertex of the lattice is 
given by combinatorial-$R$ matrix of ${U'}_q(A_n^{(1)})$, and
the states on each link are represented as the $M$-fold symmetric tensor
 representation $B_M$. 
For the simplest BBS in which only one kind of balls exist and all the
box capacity is one, the lattice model has the space of horizontal links
${B_1}^{\otimes \infty} \equiv \cdots \otimes B_1 \otimes B_1 \otimes \cdots $,
and that of the vertical links 
${B_{\infty}}^{\otimes \infty} \equiv \cdots \otimes B_{\infty} \otimes B_{\infty} \otimes \cdots $.
Here $B_{\infty}$ is understood as $B_{\infty} = B_N$ ($N \gg 1$).
Precisely speaking, $N$ can be any positive integer which is greater than
the number of the balls in the BBS.
The combinatorial $R$ matrix of ${U'}_q(A_{1}^{(1)})$ gives an isomorphism 
$\displaystyle B_{\infty}\otimes B_{1} \mapto B_1 \otimes B_{\infty}$.
The initial condition of the BBS corresponds to the initial state of the 
horizontal links of the lattice model.
Since the number of balls is finite, the initial state 
$\cdots \otimes \ket{u_{n-1}^{t=0}} \otimes \ket{u_n^{t=0}}
\otimes \ket{u_{n+1}^{t=0}} \otimes \cdots$ $\left( \in \right. $ 
 $\left. {B_1}^{\otimes \infty} \equiv \cdots \otimes B_1 \otimes B_1 \otimes
 \cdots \right)$ satisfies the condition:
$$
 \ket{u_n^{t=0}} =\ket{0} \qquad \mbox{\rm for $|n| \gg 1$},
$$
where $\ket{0}$ denotes the highest weight vector of $B_1$.
(The basis of $B_1$ will be denoted by $\ket{0}$ and $\ket{1}$, and 
$\ket{0}$ corresponds to an empty box and $\ket{1}$ corresponds 
to a box with a ball.)
The boundary condition for the vertical links is expressed as
 $\ket{v_n^{t}} = \ket{\{0\}}$ for $(|n| \gg 1)$, 
where $\ket{\{0\}}$ is the highest weight vector of $B_{\infty}$.
In general, we can replace $B_1$ with $B_{\theta_n}$ and $B_{\infty}$ 
with $B_{\kappa_t}$, where $\theta_n$ is the box capacity of
$n$th box, and $\kappa_t$ is the carrier capacity of $t$th carrying 
cart\cite{HHIKTT}.
The solitonic natures of the BBS can be proved algebraically with the above
setting\cite{FOY} and the BBS can be extended to other quantum 
algebras\cite{HKT}.

The pBBS discussed in the previous sections is also reformulated as a 
combinatorial $R$ matrix lattice model with periodic boundary condition.
For the original BBS, time evolution is given by the isomorphism:
\begin{eqnarray*}
  && \mathcal{T}: \ B_\infty \otimes B_1^{\otimes N} \to B_1^{\otimes N} \otimes B_\infty \\
  && \mathcal{T}: \ \ket{\{0\}} \otimes \ket{c(t)} \mapsto \ket{c(t+1)} \otimes \ket{\{0\}}
\end{eqnarray*}
where $\ket{c(t)} \in B_1^{\otimes N}$ is the state corresponding to the BBS at time $t$.
For the pBBS, we have to take the trace of the vertical state, {\it i.e.}, 
by regarding $\mathcal{T} \in \mbox{End}_{\mbox{\footnotesize End} B_1^{\otimes N}} B_{\infty}$, 
we define the matrix $T:=\mathrm{Tr}_{B_\infty} \mathcal{T} \in \mbox{End}_{\C} B_1^{\otimes N}$, which gives a time evolution as:
\begin{eqnarray*}
  & & T: \ B_1^{\otimes N} \to B_1^{\otimes N} \\
  & & T: \ket{c(t)} \mapsto \ket{c(t+1)} .
\end{eqnarray*}
At a glance, one may think that $\ket{c(t+1)}$ becomes a linear combination 
of many of the tensor products of $B_1$ crystals, however,
one tensor products of $B_1$ crystals map to a unique tensor
product of $B_1$ and the resultant state exactly corresponds to the state 
of the pBBS at time step $t+1$. 
Even for the pBBS with $M$ ($M \ge 2$) kinds of balls and various box 
capacities, the above lattice model is also well defined as far as the 
dimension of the vertical crystal is large enough, that is, $\kappa_t \gg 1$
for the vertical crystal $B_{\kappa_t}$ of ${U'}_q(A_{M-1}^{(1)})$.
We will show a proof of this fact for the case with one kind of balls.
Since the evolution rule for $M$ kinds of balls is decomposed
into $M$ steps as far as $\kappa_t \gg 1$, and only one kind of balls
 are moved at each step according to the same evolution rule, the proof is
also true for the case with many kinds of balls.
When $\kappa$ is small, however, the above construction will not give 
a unique tensor product and will not define an evolution rule of the pBBS.

Since we treat only one kind of balls, the states are represented by
${U'}_q(A_1^{(1)})$ crystal.
First we consider the isomorphism $B_{\kappa} \otimes B_{\theta} \simeq 
B_{\theta} \otimes B_{\kappa}$ given by the combinatorial $R$ matrix.
A state $b$ in $B_{\kappa}$ is usually denoted by a single raw semistandard
Young tableaux of length $\kappa$ on letters $1$ and $2$.
Instead we denote $b=(y, \kappa-y)$ where $y$ is the number of $1$ in the
Young tableaux.
For $(y, \kappa-y) \otimes (x, \theta-x) \simeq (x', \theta-x') \otimes 
(y', \kappa-y')$, we have the relation\cite{NY}
\begin{eqnarray}
x'&=&y-\min[\kappa, x+y]+\min[\theta,x+y]\\
y'&=&x+\min[\kappa, x+y]-\min[\theta,x+y].
\end{eqnarray}
For $\kappa > \theta$, the relation is explicitly written as
\begin{eqnarray}
y'&= &\left\{
\begin{array}{ll}
x \quad &( x+y \le \theta)\\
2x+y-\theta \quad &(\theta < x+y \le \kappa)\\
x+\kappa-\theta \quad &(\kappa < x+y)
\end{array}
\right. \\
x' &= &x+y-y'
\label{Eq.xxyy}
\end{eqnarray}
Now let $\theta_n$ $(n=1,2,\ldots,N)$ be the capacity of $n$th box, 
and $\kappa_t$
be the capacity of the carrying cart at time step $t$.
The state at time step $t$ is given by $\ket{c(t)} \in
B_{\theta_1} \otimes B_{\theta_2} \otimes \cdots \otimes B_{\theta_N}$.
Since $B_{\theta_n}$ is a ${U'}_q(A_{1}^{(1)})$ crystal, a vector $b_n \in
B_{\theta_n}$ is represented as $b_n=(x_n, \theta_n-x_n)$, where $x_n$ 
corresponds to the number of the balls in the $n$th box.
We denote a state $b_1\otimes b_2 \otimes \cdots \otimes b_N$ by
 $[x_1, x_2, \ldots, x_N]$ for $b_i=(x_i, \theta_i-x_i)$ ($i=1, 2, \ldots, N$).
The combinatorial $R$ matrix of ${U'}_q(A_{1}^{(1)})$ gives the isomorphism
 $\mathcal{T}$:
\begin{eqnarray*}
\mathcal{T}:&&B_{\kappa_t} \otimes \left( B_{\theta_1} \otimes B_{\theta_2} \otimes \cdots \otimes B_{\theta_N} \right)
\simeq \left( B_{\theta_1} \otimes B_{\theta_2} \otimes \cdots \otimes B_{\theta_N} \right) \otimes B_{\kappa_t}\\
&&\left( [y_0] \otimes [x_1, x_2, \ldots, x_N] \right. 
\left.\simeq [x'_1, x'_2, \ldots, x'_N] \otimes [y'_0] \right)
\end{eqnarray*}
From Eq.~(\ref{Eq.xxyy}), we have the following recurrence equations:
\begin{eqnarray}
y_n &= &F(y_{n-1}; x_n, \theta_n) \nonumber \\
&:= &\left\{
\begin{array}{ll}
x_n \quad &( x_n+y_{n-1} \le \theta_n)\\
2x_n+y_{n-1}-\theta_n \quad &(\theta_n < x_n+y_{n-1} \le \kappa_t)\\
x_n+\kappa_t-\theta_n \quad &(\kappa_t < x_n+y_{n-1})
\end{array}
\right. \nonumber \\
x'_{n} &= &\left\{
\begin{array}{ll}
y_{n-1} \quad &( x_n+y_{n-1} \le \theta_n)\\
\theta_n - x_n \quad &( \theta_n < x_n+y_{n-1} \le \kappa_t)\\
y_n-\kappa_n+\theta_n \quad &(\kappa_t < x_n+y_{n-1})
\end{array}
\right. 
\label{Eq.linear}
\\
&&\quad (n=1,2,\ldots, N) \nonumber \\
y'_0&= &y_{N}. \nonumber
\end{eqnarray}
We see that the function $F(y; x, \theta)$ is a piecewise linear and 
monotonically increasing function of $y$ which satisfies
$F(y+1; x, \theta)-F(y; x, \theta)=0$ or $1$ and 
$0 \le F(y; x, \theta) \le \kappa_t$.
Since $y'_0$ is a function of $y_0$, we denote it by
$y'_0 = F_N(y_0; \{x_i\}, \{\theta_i\}) := 
\underbrace{\left(F \circ F \circ \cdots \circ F\right)}_{\scriptsize \mbox{$N$ times}}(y_0)$.
The function $F_N$ is also monotonically increasing piecewise linear function,
and $ F_N(y_0+1; \{x_i\}, \{\theta_i\}) - F_N(y_0; \{x_i\},\\
 \{\theta_i\})=0$ or $1$ and $0 \le F_N \le \kappa_t$.
Thus, for $0 \le y_0 \le \kappa_t$, there is one and only one integer $y^*$
or one and only one finite interval $[y_*, y^*]$ $( y_*, y^* \in \Z)$
where the identity $y_0 =F_N(y_0 ;\{x_i\},\{\theta_i\})$ holds.
Furthermore, from Eqs.~(\ref{Eq.linear}), we find $\{x'_n\}$ do not vary
for $y_* \le y_0 \le y^*$.
Therefore we conclude that, for given $\{x_i\}$ and $\{\theta_i\}$,
there is at least one $y_0$ $(0 \le y_0 \le \kappa_t)$ at which 
$y'_0$ is equal to $y_0$ and that $\{x'_i\}$ are uniquely determined and
have the same values for $y_0$ which satisfies $y'_0=y_0$.
The above conclusion means that $T:=\mathrm{Tr}_{B_{\kappa_t}} \mathcal{T}
\in \mathrm{End}B_{\theta_1}\otimes\cdots\otimes B_{\theta_N}$
maps a state $b_1 \otimes \cdots \otimes b_N$ to the state which is also
described by one tensor product of crystal basis.
We summarize the above statement as a Theorem.

\begin{Theorem}
If $\kappa$ is greater than any $\theta_i$ ($i=1, 2, \ldots,N)$, 
the map $T:=\mathrm{Tr}_{B_{\kappa}} \mathcal{T}$ $\in \mathrm{End} 
B_{\theta_1} \otimes B_{\theta_2} \otimes \cdots \otimes B_{\theta_N}$
sends a tensor product of crystal basis to a unique tensor product 
of crystal basis of \/ ${U'}_q(A_1^{(1)})$.
Furthermore, for sufficiently large $\kappa$, the statement holds for
${U'}_q(A_M^{(1)})$ with arbitrary positive integer $M$.
\end{Theorem}

From the construction of the map, it is clear that the pBBS discussed 
in the previous section corresponds to the case $\theta_i=1$ $(i=1,2,
\ldots, N)$, and we use this map to construct the pBBS with arbitrary
box capacities and ball species.

\subsection{pBBS as $A_{N-1}^{(1)}$ crystal chains}

\begin{figure}[!b]
  \begin{center}
    \scalebox{0.35}{\includegraphics{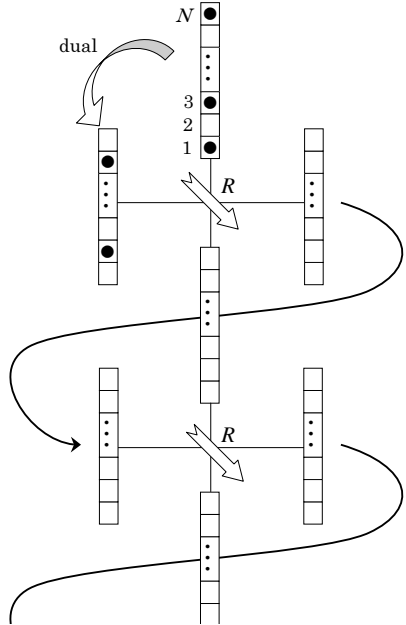}}
  \end{center}
  \caption{Twisted chains of crystal $A_{N-1}^{(1)}$ and pBBS.}
  \label{fig:slN}
\end{figure}

When there are $M$ balls of the same kind and $N$ boxes, we can also 
reformulate the pBBS in terms of the combinatorial $R$ matrix 
of ${U'}_q(A_{N-1}^{(1)})$ and symmetric tensor product $B_M$ and $B_{M'}$
where $\displaystyle M':=\sum_{i=1}^N \theta_i -M$
 (Fig.~\ref{fig:slN}).
 
A crystal $b \in B_{M}$ can be denoted by $b=(x_1,x_2, \ldots, x_N)$ with 
$0 \le x_i \le \theta_i$, $\sum_{i=1}^{N}x_i= M$. 
We associate a state of pBBS with the crystal $b$ in which $x_i$ is the number
of balls in the $i$th box of the state.
For the crystal $b$, we define the dual crystal $\bar{b}= (\bar{x}_1, 
\bar{x}_2, \ldots, \bar{x}_N) \in B_{M'}$, where $\bar{x}_i = \theta_i-x_i$ 
$(i=1,2,\cdots,N)$.
Then the crystal $b' \in B_{M}$ associated with the state at time $t+1$ is
given by the combinatorial $R$ matrix which gives the isomorphism
 $ B_{M'} \otimes B_{M} \simeq B_{M} \otimes 
B_{M'}$ as
\begin{equation}
R: \quad \bar{b} \otimes b \mapto b' \otimes \bar{b'}.
\label{Evolution.R}
\end{equation}
From ref.~\cite{NY}, we see that this gives the same time evolution
of the pBBS discussed above.
As is shown in Fig.~(\ref{fig:slN}), the time evolution is described in
two twisted chains of $B_M$ and $B_{M'}$.
Note that by changing the crystal $b$ and/or the $\bar{b}$ 
with another crystal (say $B$ type crystal), we obtain other types of pBBS 
with a time evolution rule given by the isomorphism $R$.
We may find interesting features in these CAs, however, the investigation 
of these pBBSs is a future problem.

The isomorphism (\ref{Evolution.R}) has been shown to be expressed 
as an ultradiscrete KP equation (Eqs.~(22) and 
(23) in ref.~\cite{HHIKTT}), which serves another reason why we claim the 
pBBS is an integrable CA.
Here we do not repeat the results in ref.\cite{HHIKTT}, but show a similar
formulae to (\ref{eq:req1}).
The space of the states, however, is no longer a finite field but
 $\Z^N$.
For $X=(x_1,x_2, \ldots, x_N),\ Y=(y_1, y_2, \ldots, y_N) \in \Z^N$, 
we define $\max$ and $\min$: $\Z^N \times \Z^N
 \to \Z^N$ as
\begin{eqnarray*}
\left(\min[X,Y]\right)_i &= &\min[x_i,y_i]\\
\left(\max[X,Y]\right)_i &= &\max[x_i,y_i].
\end{eqnarray*}
We also define the rotate shift to the right $S$: 
$\Z^N \to \Z^N$ as
\[
SX=(x_N, x_1, x_2, \ldots, x_{N-1}).
\]
Let $\theta_i (\in \Z_{>0})$ be the capacity of $i$th box and
$x_i(t)$ ($0 \le x_i(t) \le \theta_i$) be the number of balls in 
$i$th box at time step $t$.
We denote the state of the pBBS at $t$ by $X(t) := (x_1(t), x_2(t), 
\ldots, x_N(t))$.
The state at $t+1$, $X(t+1)$, is obtained from the following Theorem:

\begin{Theorem}
\label{Thm:CB}
Let $A^{(0)}=X(t)$ and $B^{(0)}=S X(t)$.
We define $A^{(n)}$ and $B^{(n)}$ ($n=1,2,\ldots)$ by the recurrence equations:
\begin{eqnarray}
  \left\{
    \begin{array}{l}
      A^{(n+1)} := \min \left[A^{(n)} + B^{(n)}, \boldsymbol{\theta} \right]\\
      B^{(n+1)} := S \max \left[ A^{(n)} + B^{(n)} -\boldsymbol{\theta}, 
      \mbox{\boldmath{$0$}}\right] \\
    \end{array}
  \right.,
  \label{eq:reqZ}
\end{eqnarray}
where $\boldsymbol{\theta} := (\theta_1, \theta_2, \ldots, \theta_N)$.
Then we obtain 
\begin{equation}
  X(t+1) = A^{(N-1)} - X(t),\ 
  B^{(N-1)}= \mbox{\boldmath{$0$}} .
  \label{eq:reqfinZ}
\end{equation}

\end{Theorem}

\begin{Proof}
The proof of Theorem \ref{Thm:CB} is similar to 
that of the Theorem 1 and we simply show its outline.
We define $D^{(i)}$ $(i=1,2, \ldots, N)$ as in the proof of Theorem 1.
Then we have the decomposition:
\begin{equation}
  X(t)=\displaystyle \sum_{i=1}^{N-1} D^{(i)}, \ \ 
  X(t+1)=\displaystyle \sum_{i=1}^{N-1} S^i D^{(i)} .
\end{equation}
From the recurrence equations (\ref{eq:reqZ}) and the properties of the
similar diagram of evolution patterns, we can inductively show
\begin{eqnarray}
  \left\{
    \begin{array}{l}
      A^{(k)} = A^{(k-1)} + S^{k} D^{(k)} \\
      S^{-k-1} B^{(k)} = S^{-k}B^{(k-1)} - D^{(k)} \\
      S^{k+1} D^{(k+1)} = \min\left[\boldsymbol{\theta} -A^{(k)}, B^{(k)} 
      \right]
    \end{array}
  \right.,
  \label{eq:RecReal}
\end{eqnarray}
and we have
\begin{eqnarray}
A^{(n)}&= &X(t) +\sum_{i=1}^{n} S^i D^{(i)} ,\\
B^{(n)}&= &S^{n+1}\left( \sum_{i=n+1}^{N-1} D^{(i)} \right) ,
\end{eqnarray}
which completes the proof.

\end{Proof}

\section{Summary}

In this paper, we introduced the pBBS which is an extension
of the original BBS to that with a periodic boundary condition. 
We showed that the evolution rule of the pBBS is given by a recurrence
Boolean formula
which is regarded as an ultradiscretized algorithm of the calculation of 
$2N$th root of a given number.
We also gave the conserved quantities of the recurrence formula.
The relation to the combinatorial $R$ matrix of 
$\displaystyle {U'}_q(A_{N-1}^{(1)})$ 
was clarified,
and generalization of the pBBS with the symmetric tensor product
 representaion of
$\displaystyle {U'}_q(A_{N-1}^{(1)})$ crystals was discussed.

Since the pBBS takes only finite number of states, it has a fundamental
 cycle.
Determination of this fundamental cycle is one of the future problems.
In addition, an integrable equation usually has quasi-periodic solutions
given by Theta functions.
A state of the pBBS is expected to be obtained from the quasi-periodic
 solution through ultradiscretization.
To obtain rigorous expression of the solutions to the pBBS through 
ultradiscretization is another future problem.

\end{document}